# Phase Equilibria and Crystal Growth for LiREF$_4$ Scheelite Crystals

Detlef Klimm[1], Ivanildo A. dos Santos[2], Izilda M. Ranieri[2], and Sonia L. Baldochi[2]

[1]Leibniz Institute for Crystal Growth, Max-Born-Str. 2, 12489 Berlin, Germany

[2]Center for Lasers and Applications, IPEN, Butantã 05422-970, São Paulo, SP, Brazil

**ABSTRACT**

The scheelite type laser crystals LiREF$_4$ melt congruently only for RE being one of the elements Er, Tm, Yb, Lu, or possibly Y, respectively. For RE = Eu, Gd, Tb, Dy, or Ho the corresponding scheelites undergo a peritectic melting under the formation of the corresponding rare earth fluoride. The melting behavior of LiREF$_4$ mixed crystals with two or more RE is not yet known well. If RE is a mixture of Gd and Lu, Gd rich solid solutions melt peritectically under formation of (Gd,Lu)F$_3$ and Lu rich solid solutions melt directly without formation of other solid phases.

**INTRODUCTION**

Most of the binary systems LiF–REF$_3$, where RE is one of the rare earth elements from lanthanum to lutetium, or yttrium, respectively, were described by Thoma [1]. For the smaller rare earth ions, starting from RE = Eu, the systems contain one intermediate phase LiREF$_4$ that crystallizes at ambient conditions in the scheelite structure. At high pressure (10.7 GPa in the case of LiLuF$_4$ [2]) a reversible phase transformation to a monoclinic phase is observed for the smaller RE, contrary the scheelites decompose under formation of RE rich phases and LiF for the larger RE (beyond 11 GPa in the case of LiGdF$_4$ [3]). Solid state synthesis and lattice parameters for these compounds were reported by Keller and Schmutz [4]. The thermal stability of the scheelites under ambient pressure is larger for the smaller RE$^{3+}$. LiEuF$_4$ is stable up to 690°C where it decomposes peritectically under the formation of β-EuF$_3$. Peritectic decomposition to a melt and the low-$T$ (β-) phase of the REF$_3$ is also observed for RE = Gd, Tb, Dy, Ho, and Y; but the peritectic point shifts in this sequence closer to the solidus of the LiREF$_4$ phase. LiErF$_4$ is controversially reported to melt incongruently [5] or congruently [1]. If one assumes a smooth variation of the melting temperatures $T_\text{f}$ for neighboring RE, congruent melting seems more realistic, since LiTmF$_4$, LiYbF$_4$, and LiLuF$_4$ melt congruently [1]. Recently from a DFT evaluation structural parameters $a_0$, $c_0$, elastic stiffness coefficients $c_{ij}$, and thermodynamic parameters were derived for most of the LiREF$_4$ [6]. It was concluded that the scheelite structure should be stable also for the large RE from Pm to Ce, but the energies for the LiREF$_4$ from Gd to Lu are more favorable, which is in reasonable agreement with the experimental phase diagrams [1].

The LiREF$_4$ scheelites attract not only academic interest as some of them are interesting laser host materials where the host RE$^{3+}$ can be replaced by another rare earth laser ion RE'$^{3+}$. LiYF$_4$ (YLF), often doped with Nd, is a good example that is offered by several suppliers commercially. YLF crystals can be grown from melts with slight LiF excess (molar fraction $x_\text{Y}$=0.48) e.g. by the Czochralski or Bridgman technique [7,8]. The fluoride crystals are sensitive against hydrolysis during growth which can be suppressed e.g. by an atmosphere containing CF$_4$ or even HF during the Czochralski process [9]. This paper reports new and reviews recent results on phase relations and crystal growth within binary $LiREF_4$–$LiRE'F_4$ systems.

**EXPERIMENT**

LiF (Aldrich, 99.9% = 3N purity) was zone melted for purification in a platinum boat (cross section 1 cm$^2$, length 20 cm) inside a platinum tube that was rinsed by a HF/Ar mixture. $REF_3$ (RE = Gd, Lu, or Y) were prepared from commercial $RE_2O_3$ (typically 3N–5N purity) by hydrofluorination at 850°C in the same apparatus. This process is described in detail elsewhere [10]. Conversion rates >99.9% of the theoretical value calculated for the reaction $RE_2O_3 + 6HF \rightarrow 2REF_3 + 3H_2O$ were measured by comparing the masses prior to and after the hydrofluorination process. The samples were mainly prepared by melting and slow cooling (15 K/hour) under HF/Ar. Only for $LiLuF_4$ and $LiGdF_4$ pieces from Czochralski grown single crystals were used. It will be shown later that the melting behavior of most $LiREF_4$ mixed crystals can only be described, if the phase relations of $REF_3$ mixed crystals are known and consequently some of these system were studied first.

Three different NETZSCH simultaneous DTA/TG or DSC/TG equipments were used for thermal analysis: The systems $REF_3$–$RE'F_3$ and $LiREF_4$–$LiRE'F_4$ were investigated using a STA 449C (Pt/Rh furnace, DSC sample holder, graphite or Pt crucibles) and a STA 409CD (graphite furnace, DTA sample holder, graphite crucibles) or a STA 409 PC/PG (SiC furnace, DSC sample holder, Pt crucibles). The crucibles were covered with lids. After evacuation of the thermal analyzers the samples were measured in flowing Ar (99.999% purity) with a heating rate of 10 K/min. Such procedure could avoid hydrolysis of the sensitive fluorides almost completely.

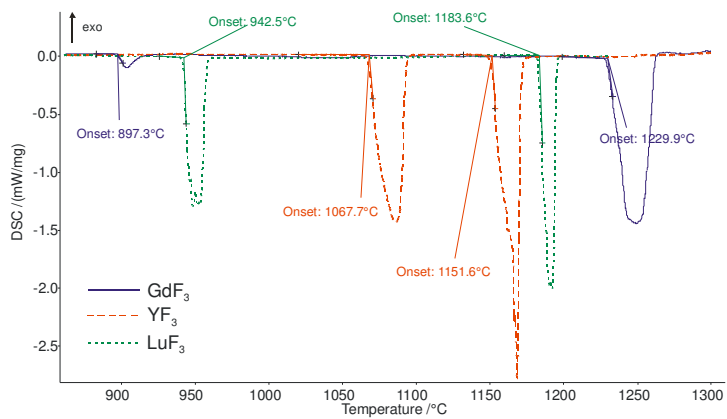

**Fig. 1: DSC curves (2$^{nd}$ heating run, original data without smoothing) of pure $REF_3$ showing subsequently peaks from the phase transformation ($T_t$) and from melting ($T_f$).**

Usually only heating curves were used for the construction of phase diagrams, as cooling curves did often show supercooling. A typical example for $GdF_3$, $YF_3$, and $LuF_3$ is shown in Fig. 1. Powder samples (≈10 µm grains) with masses around 20 mg as obtained from the hydrofluorination process were used for these measurements. The 2$^{nd}$ heating curves that are shown in Fig. 1 do not differ remarkably from the 1$^{st}$ heating curves that were obtained with the initial powders. After the DSC runs the molten and solidified samples formed one drop on the bottom of the crucible. The small mass loss (not shown in the figure, but never exceeding 4%; <1% are typical) during these measurements and mainly well shaped sharp peaks prove that the samples and the gas flow in the thermal analyzer were free from oxygen and humidity. It is obvious that all three REF$_3$ show two subsequent thermal effects: A first endothermal peak at $T_t$ indicates a first order solid state phase transformation and a second endothermal peak at $T_f$ indicates melting. Similar measurements were performed with binary mixtures GdF$_3$–YF$_3$ [11] and GdF$_3$–LuF$_3$ [12].

As expected, both peaks become broader for intermediate compositions because phase transformation as well as melting take place in a temperature interval instead of a fixed $T$ for solid solutions. In the case of GdF$_3$–YF$_3$, it was found that the onset of the phase transformation peak changes almost continuously from $T_t \approx 900°C$ (GdF$_3$) to $T_t \approx 1066°C$ (YF$_3$), passing a weak maximum around 10% GdF$_3$ where $T_t$ reaches 1080°C. (All concentrations throughout the text are given in mol-%). The melting point changes also continuously from $T_f \approx 1230…1250°C$ (GdF$_3$) to $T_t \approx 1130…1150°C$ (YF$_3$), passing a weak minimum around 25% GdF$_3$ where $T_f$ reaches 1120°C [11].

Conversely, GdF$_3$–LuF$_3$ mixtures show a different behavior: Here $T_t$ rises from both sides ($\approx 900°C$ or $\approx 945°C$, respectively) and reaches a constant value $T_t \approx 1051°C$ for intermediate compositions from 20 to 60% LuF$_3$. Instead, $T_f$ drops from both sides (1250°C or 1182°C, respectively) to a constant eutectic temperature $T_{eut} = 1092°C$ that can be observed for compositions between 20 and 70% LuF$_3$. Moreover, nine GdF$_3$–LuF$_3$ samples with different composition were pulverized and lattice constants were measured using a Bruker AXS diffractometer. The diffraction patterns could be fitted satisfactory to the *Pnma* space group and no parasitic peaks indicating other phases were found [12].

As written above, most LiF–REF$_3$ phase diagrams are already published [1], and only the system LiF–GdF$_3$ was re-determined recently [13]. For practical applications the scheelites LiREF$_4$ (usually doped with another RE′) are the most interesting phases within these systems, but the knowledge on pseudo-binary systems LiREF$_4$–LiRE′F$_4$ is scarce. DSC results on LiGdF$_4$, LiLuF$_4$, and their binary mixtures were published in a previous paper [14].

A single mixed crystal was grown from the composition Li(Lu$_{0.75}$Gd$_{0.25}$)F$_4$ by the Czochralski method under high purity CF$_4$ + Ar atmosphere, in a commercial system with automated diameter control. The crystal was pulled with a growth rate of 1 mm/hour and a rotation rate of 15 rpm for a [100] oriented boule, a seed of LiLuF$_4$ was used to achieve the crystallization process. The crystal with 50 mm in length, 15 mm in diameter and 52 g is shown in Fig. 2. The crystal is optically clear and grew well.

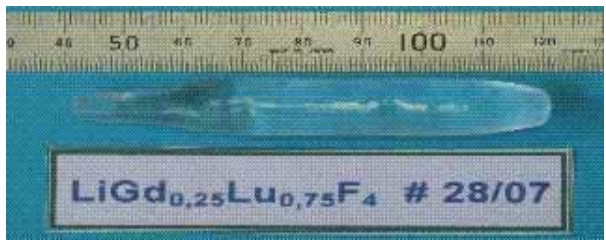

**Fig. 2: Czochralski grown scheelite mixed crystal Li(Lu$_{0.75}$Gd$_{0.25}$)F$_4$.**

## DISCUSSION

The concentration triangle LiF–GdF$_3$–LuF$_3$ shall be used as an example for the discussion of phase equilibria that are relevant for the growth of scheelite mixed crystals. The scheelites can be found on the dashed line in Fig. 3.

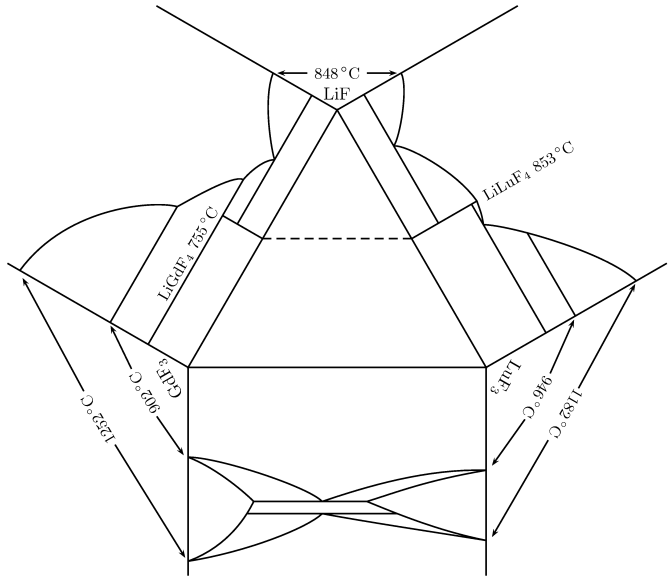

**Fig. 3: Concentration triangle LiF–GdF$_3$–LuF$_3$ with intermediate scheelites LiGdF$_4$ (incongruently melting) and LiLuF$_4$ (congruently melting)**

In should be noted that binary subsystems in phase diagrams can only be spanned between compounds with congruent melting behavior. This is a necessary, but not sufficient condition, and this condition is fulfilled only for a few RE-RE´ pairs. In Fig. 3, only LiLuF$_4$ shows congruent melting, whereas LiGdF$_4$ melts incongruently. Besides LiLuF$_4$, the LiREF$_4$ scheelites with RE = Yb, Tm, Er show congruent melting behavior too. LiYF$_4$ is arguable, as the temperature of its solidus is at least so close to the liquidus at this composition that they cannot be resolved by means of thermal analysis. All other LiREF$_4$ undergo a peritectic decomposition to a melt and remaining REF$_3$ in the low-$T$ phase, which is for all rare earth fluorides the β-YF$_3$ type. From the data that are given in Tab. 1 it becomes obvious that at last for the melting of pure LiREF$_4$ never the formation of the high-$T$ REF$_3$ phase has to be taken in account.

| RE | Eu | Gd | Tb | Dy | Ho | Y | Er | Tm | Yb | Lu |
|---|---|---|---|---|---|---|---|---|---|---|
| LiREF$_4$ melting | 690 per. | 755 per. | 787 per. | 819 per. | 799 per. | 812 con.? | 832 con. | 785 con. | 790 con. | 794 con. |
| LiREF$_4$ liquidus | 763 | 857 | 842 | 843 | 819 | ≈812 | 832 | 785 | 790 | 794 |
| REF$_3$ low-$T$ | β-YF$_3$ type | | | | | | | | | |
| $T_t$ | 765 | 878 | 998 | 1082 | 1100 | 1059 | 1101 | 1044 | 949 | 937 |
| REF$_3$ high-$T$ | tysonite type | | | | | | α-UO$_3$ type | | | |

**Tab. 1: Melting temperature and melting behavior for the known LiREF$_4$ scheelites, together with the corresponding solid state phase transformation temperatures $T_t$ of the corresponding REF$_3$ and their structure types below and above $T_t$ (temperatures in °C).**

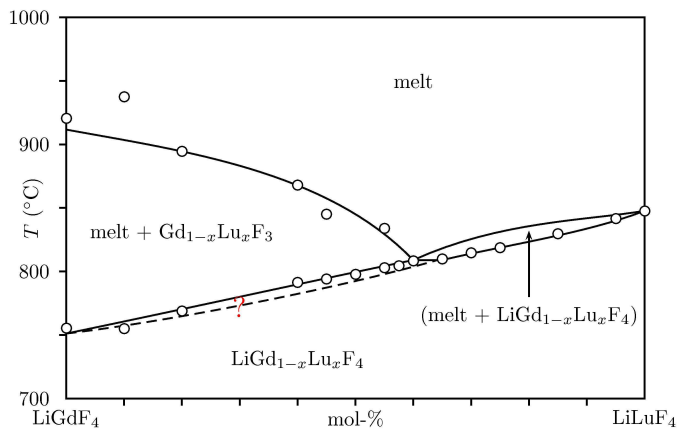

Fig. 4: The tentative scheelite section (dashed line in Fig. 3) showing peritectic decomposition scheelite → melt + (Gd,Lu)F$_3$ for LiGdF$_4$ rich compositions and incongruent melting for LiLuF$_4$ rich compositions [14].

As mentioned above, pseudo-binary phase diagrams can be constructed between the congruently melting LiREF$_4$, this means for RE, RE′ = {Lu, Yb, Tm, Er, and possibly Y}. To the author's knowledge, none of these (5×4)/2 = 10 pseudo-binary systems was measured up to now. However, the scheelite section LiGdF$_4$–LiLuF$_4$ of the ternary system LiF–GdF$_3$–LuF$_3$ was investigated recently [14], and it was found that for compositions close to LiGdF$_4$ (<60 mol-% LiLuF$_4$) the mixed crystal scheelite decomposes peritectically to (Gd,Lu)F$_3$ and melt, whereas LiLuF$_4$ rich compositions melt directly, without peritectic decomposition (Fig. 4). It should be noted, however, that a mixed crystal usually not melts congruently: Congruent melting would mean that a solid body and a liquid phase with identical chemical composition are in equilibrium. This is for mixed crystals the case only at azeotropic points where liquidus and solidus have one common maximum (e.g. SrNb$_2$O$_6$–BaNb$_2$O$_6$ [15]) or minimum (e.g. CaF$_2$–SrF$_2$ [16]), and such azeotropic point does not occur at least in Fig. 4.

## CONCLUSIONS

The melting and crystallization behavior of scheelite type LiREF$_4$ mixed crystals, where the rare earth element is partially substituted by another RE´, can in most cases only be described in terms of ternary phase diagrams. Only if both rare earth elements belong to the set {Lu, Yb, Tm, Er, and possibly Y}, it can be expected that the scheelite mixed crystals behave as binary mixtures.

The 2-phase-field "melt+scheelite" is, at least for RE, RE´ = Gd, Lu, very narrow, resulting in a segregation coefficients of both rare earth elements during crystal growth that are close to unity. Consequently, the melt growth of single crystals with good homogeneity is possible outside the composition region where peritectic decomposition of the scheelite to a melt and rare earth fluoride mixed crystal takes place. A similar narrow separation between liquidus and solidus of scheelite mixed crystal was reported by Abell et al. [17] for the system LiYF$_4$–LiErF$_4$ where both end members melt congruently and homogeneous crystals Li(Y,Er)F$_4$ without segregation could be grown. It should be noted, however, that Y$^{3+}$ and Er$^{3+}$ have almost identical ionic radii (104 or 103 pm, respectively for octahedral coordination, Shannon) compared with Gd$^{3+}$ and Lu$^{3+}$ (107.8 or 100.1 pm, respectively) — but the larger difference of the pair Gd$^{3+}$-Lu$^{3+}$ does not influence the melting behavior considerably.

**ACKNOWLEDGMENTS**

The authors express their gratitude to "Coordenação de Aperfeiçoamento de Pessoal de Nivel Superior" (CAPES) and to "Deutscher Akademischer Austauschdienst" (DAAD) for financial support.